\documentclass{article}
\usepackage{ijcai20}

\usepackage[normalem]{ulem}
\usepackage{times}
\usepackage{soul}
\usepackage{url}
\usepackage[hidelinks]{hyperref}
\usepackage[utf8]{inputenc}
\usepackage[small]{caption}
\usepackage{graphicx}
\usepackage{color}
\usepackage{amsmath}
\usepackage{amsthm}
\usepackage{booktabs}
\usepackage{algorithm}
\usepackage{algorithmic}
\usepackage{threeparttable}
\title{\textit{Privacy for Rescue}: A New Testimony Why Privacy is Vulnerable In Deep Models}

\author{
Ruiyuan Gao
\and
Ming Dun\and
Hailong Yang\and
Zhongzhi Luan\and
Depei Qian
\affiliations
Beihang University
\emails
\{gaoruiyuan, dunming0301, hailong.yang, 07680, depeiq\}@buaa.edu.cn
}

\begin{document}

\maketitle

\begin{abstract}
The huge computation demand of deep learning models and limited computation resources on the edge devices calls for the cooperation between edge device and cloud service by splitting the deep models into two halves.
However, transferring the intermediate results from the partial models between edge device and cloud service makes the user privacy vulnerable since the attacker can intercept the intermediate results and extract privacy information from them. Existing research works rely on metrics that are either impractical or insufficient to measure the effectiveness of privacy protection methods in the above scenario, especially from the aspect of a single user. In this paper, we first present a formal definition of the privacy protection problem in the edge-cloud system running DNN models.
Then, we analyze the-state-of-the-art methods and point out the drawbacks of their methods, especially the evaluation metrics such as the Mutual Information (MI). In addition, we perform several experiments to demonstrate that although existing methods perform well under MI, they are not effective enough to protect the privacy of a single user. To address the drawbacks of the evaluation metrics, we propose two new metrics that are more accurate to measure the effectiveness of privacy protection methods. Finally, we highlight several potential research directions to encourage future efforts addressing the privacy protection problem.
\end{abstract}

\section{Introduction}
Despite the powerful generalization capability of Deep Neural Networks (DNNs), their computation demands are tremendous. Especially in the field of edge computing, the edge devices are commonly utilized by deep learning applications in combination with online services (or cloud), due to the limited computing ability. In such case, users need to provide the data (e.g., images) generated at the edge devices (e.g., camera) to the service providers in order to get their responses (e.g., objects in an image). Although the collaboration between cloud services and edge devices boosts the development of deep learning applications in innovative ways, protecting user's privacy in such scenario becomes a major concern, which could lead to severe privacy breach if not paying enough attention.

On one hand, sending user's raw data from edge device to cloud is the most risky way, because the data can be intercepted during various stages, such as software stack~\cite{Information_Leakage_in_Third-party_Compute_Clouds}, and network communication~\cite{Security_Hazards}.
On the other hand, if all DNN computation is performed on edge device, there is no need to worry about privacy breach. However, in practice this is infeasible because the computation demands of deep models are prohibitive for edge devices. Although traditional cryptographic techniques such as multiparty execution~\cite{Multiparty_Computation} and homomorphic encryption~\cite{homomorphic_encryption} are applied to DNNs for protecting privacy, the encryption and decryption operations themselves require too much computation, which is not affordable for edge devices.

\begin{figure}[t]
    \setlength{\abovecaptionskip}{1px}
    \begin{center}
       \includegraphics[width=3.28in]{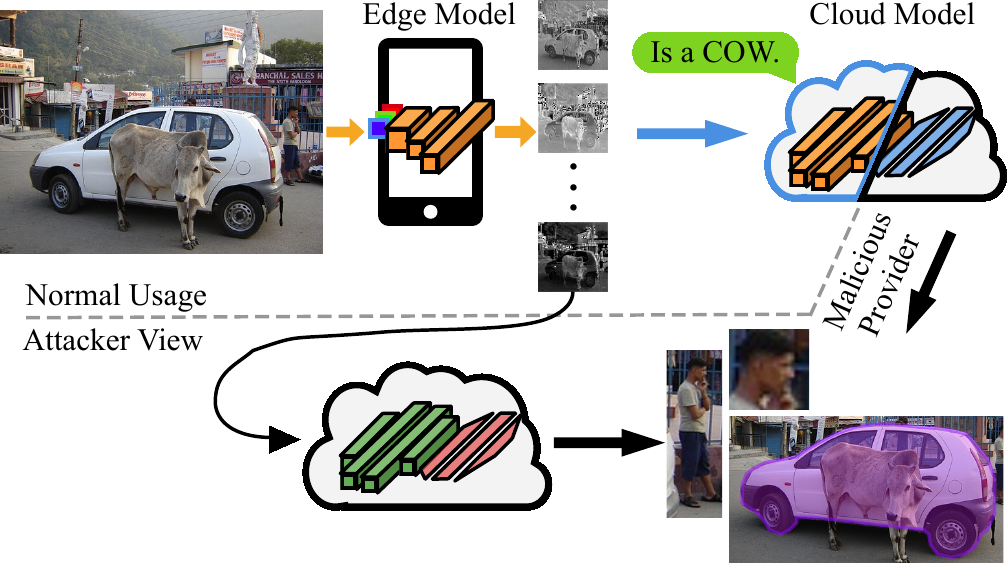}
    \end{center}
    \caption{An illustrative example for collaboration between edge device and cloud service. In this case, the user may suffer from privacy breach if the intermediate results are intercepted and combined with another deep model by the malicious provider.}
    \label{fig:overview}
 \end{figure}

In practice, a compromise is reached between the load on edge device and privacy preservation~\cite{DPFE}. A commonly adopted approach is to cut a deep model into two parts and run each part on the edge device and cloud service respectively, as shown in Fig~\ref{fig:overview}. The partial model running on the edge device (\textit{edge model}) contains the first $m$ layers, which takes user raw data as input and generates an intermediate result as output. The other partial model running on the cloud service (\textit{cloud model}) contains the rest $n-m$ layers, which takes intermediate result from the \textit{edge model} as input and generates the final result. To protect the privacy of the intermediate result, techniques such as applying noise to the intermediate~\cite{Shredder} are proposed, which reduces the Shanon’s Mutual Information (MI).

However, the research on privacy protection regarding the scenario of synergy between edge and cloud running DNN application is still preliminary and requires more research efforts. In this paper, we consider privacy attack as shown in Fig.~\ref{fig:overview}, which could come from both network hacker and malicious service provider.
Even though attackers cannot recover the raw user data, they can reveal private information from the intermediate results~\cite{TransferLearning}, which is far beyond the task that user intends to do. Existing approaches~\cite{DPFE,Shredder} prevent privacy breach by cutting deep models at appropriate layers between edge and cloud, and then applying noise or adjusting feature extraction to the intermediate results. However, existing metrics (e.g., MI) are not good indicators for measuring the effectiveness of privacy protection, especially from a single user aspect. For example, Shredder~\cite{Shredder} uses MI to represent the information loss at the scope of entire database rather than pairwise images. Therefore, the value of MI cannot be used to measure the amount of private breach for a single user.



To clearly articulate the scenario and the risk of privacy breach, we take Fig~\ref{fig:overview} for an example.
A user takes a photo and wants to identify the breed of the cow. However, the photo also accidentally contains a person, which is irrelevant to the user's task.
Normally, the service provider only performs the designed task and returns the answer to the user. However, a malicious provider may take the intermediate results sent from the user device and feed them to another pre-trained cloud model that is used to identify gender, age, and other private information of the person in the user image.
The above privacy breach still happens even if the intermediate results are applied with noise and thus exhibit lower MI~\cite{Shredder}, which we will demonstrate in the following experiments.


This paper provides a new testimony that the privacy of current edge-cloud system running deep models is vulnerable. We advocate for more research attention in such area to promote the advance of privacy protection that is far from satisfactory. Specifically, we articulate the edge-cloud scenario running deep neural models where the privacy breach could happen. In addition, we formulate the privacy protection problem and propose a new metric to measure the effectiveness of privacy protection. Moreover, we perform a serial of experiments to demonstrate that none of the existing approaches can protect the privacy perfectly in such scenario. We hope these experiment results can layout a solid foundation for further research efforts. Finally, we discuss the potential key points that need to be addressed in future studies in order to protect privacy for deep neural models running on the edge-cloud system.

\section{Related Work}

Privacy has always been a major problem for IT service providers~\cite{Privacy-Preserving_Deep_Learning}.
Ever since the realization of applying DNN models in various application domains, researchers have been thinking about how to provide privacy protection for DNN models.

Recently, many research work focuses on preserving privacy for the training procedure of DNN.
By applying a privacy-preserving strategy with federated learning, \cite{privacy-preserving} is able to provide increasing privacy protection for the training data provider.
Differential Privacy (DP)~\cite{Differential_Privacy} is also a competitive theory for privacy protection.
Benefit from its theoretical foundation, local DP has been broadly used when training a new model~\cite{Differential_Privacy}.
However, the main objective of above strategies is privacy protection for the training data provider, and their applicability at inference stage to protect privacy for a single user remains questionable.

Encryption and cryptographic techniques are naturally considered to protect privacy in the scenario of edge-cloud system running deep neural models.
Secure Multiparty Computation (SMC)~\cite{Multiparty_Computation} provides a paradigm to keep security using cloud computing at the absence of a trustworthy service provider.
\cite{homomorphic_encryption} and \cite{cryptonets} expand the traditional encryption techniques to implement SMC.
They allow cloud computing based on encrypted data.
As a result, all data sent by edge device is encrypted, and thus safe for computation and communication.
However, the above methods require a large amount of computation for encryption and decryption operations, especially when the DNN models are used to process a large volume of data. Therefore, such intrinsic drawback prevents their widely adoption in practice.

Shredder~\cite{Shredder} and DPFE~\cite{DPFE} are the most recent research works that are designed for the scenario of edge-cloud system and focus on protecting the user privacy during inference.
DPFE based on the assumption that the end user knows what kind of information they would like to protect.
However, in practice, not all end users know what information they should pay special attention to. Even worse, some users are unaware of privacy breach at all.
As for Shredder, the metric MI used for evaluating the effectiveness of privacy protection cannot truthfully reflect the privacy preservation for a single user. In addition, they assume that the attacker will not re-train the cloud model after acquiring the intermediate results, which can easily invalidate their approach of applying noise to the intermediate results for privacy protection.

\section{Understanding Privacy Breach}

To understand the privacy breach, we need to clearly define the scenario for privacy protection and how to evaluate the effectiveness of a given method.
In the section, we first present the formulated description about the scenario where the edge device cooperates with service provider running DNN models to serve user's request, and define the privacy that needs to be protected (Sec.~\ref{sec:Privacy for End User}).
After that, we point out the major drawbacks in the hypothesis or evaluation settings that makes existing approaches ineffective to protect privacy (Sec.~\ref{sec:Problem in Existing Paradigm}) in our scenario. Finally, we propose our testimony for a new set of metrics to evaluate the effectiveness of privacy protection methods for a single user (Sec.~\ref{sec:A New Testimony}). 

\subsection{Privacy for End User}\label{sec:Privacy for End User}
To deploy a pre-trained deep model $model=f(\mathbf{x};\theta)$, where $\mathbf{x}$ is input and $\theta$ is parameter, in edge-cloud system, common practice is to cut it into two parts.
$model_{l}=f_{l}(x;\theta_{l})$ stands for the first several layers of the model that are computed on edge device, whereas $model_{r}=f_{r}(x;\theta_{r})$ stands for the rest layers that are computed on cloud service. Thus, we can describe the above deployment using Eq.~\ref{eqn:1} and Eq.~\ref{eqn:2}, where $\mathbf{a}$ is the intermediate result that needs to be sent to the cloud service, and $\mathbf{y}$ is the final result.

\begin{eqnarray}
    &&\mathbf{y}=f(\mathbf{x};\theta)=f_{r}(\mathbf{a};\theta_{r})\label{eqn:1}\\
    &&\mathbf{a}=f_{l}(\mathbf{x};\theta_{l})\label{eqn:2}
\end{eqnarray}

As illustrated in Fig.~\ref{fig:overview}, in this paper, we consider the privacy breach when the intermediate results are sent from the edge device, which means we trust the model $model_{l}$ running on the edge device. However, the intermediate results could be intercepted by either network hackers or malicious providers and fed to another attacker model ($model_{a}$) that deviates from the original user task.

\newtheorem{assumption}{Assumption}
\begin{assumption}\label{Ass:question define}
The privacy breach happens when the intermediate result $\mathbf{a}$ is applied to a model that is different from $model_{r}$, deviating from its original task.
\end{assumption}

Based on Assumption~\ref{Ass:question define}, we focus on strategies that prevent applying $\mathbf{a}$ to $model_{a}$ that can reveal the privacy other rather the original user task.
The difficulty for privacy protection is that, after obtaining $\mathbf{a}$, the attacker can apply arbitrary post-processing in order to breach the privacy. Although recovering $\mathbf{x}$ through $\mathbf{a}$ seems a difficult task, the attacker can easily extract a lot of useful information through $\mathbf{a}$, which has already been proved by~\cite{DPFE}. 

To further ease the discussion, we also formulate the attacker behavior as shown in Eq.~\ref{equ:privacy digging}.
As stated in Assumption~\ref{Ass:question define}, after an attacker obtains the intermediate results $\mathbf{a}$, the attacker model, $model_{a}=f_{a}(x;\theta_{a})$ is applied to $\mathbf{a}$ for disclosing the privacy information ($\mathbf{y}_{p}$).

\begin{equation}\label{equ:privacy digging}
    \mathbf{y}_{p}=f_{a}(\mathbf{a};\theta_{a})
\end{equation}

To counter the attack, privacy protection strategy as shown in Eq.~\ref{equ:convert function} is applied, which strengthens $model_{l}$ (e.g., adding noise to $\mathbf{a}$ as shown in Eq.~\ref{equ:noise-a}), in order to reduce the accuracy of $\mathbf{y}_{p}$, while deteriorating the accuracy of $\mathbf{y}$ as little as possible, as shown in Eq.~\ref{equ:eq5} and Eq.~\ref{equ:eq6}.

\begin{eqnarray}
    f^{'}_{l}(\cdot;\theta^{'}_{l})&=&g(f_{l},\theta_{l})\label{equ:convert function}\\
    \mathbf{a}^{'}&=&f^{'}_{l}(\mathbf{x};\theta^{'}_{l})\label{equ:noise-a}\\
    \mathbf{y}&=&f_{r}(\mathbf{a}^{'};\theta_{r})\label{equ:eq5}\\
    \mathbf{y}_{p} &\neq& f^{'}_{a}(\mathbf{a}^{'};\theta^{'}_{a}), \forall f^{'}_{a}(\cdot;\theta^{'}_{a})\label{equ:eq6}
\end{eqnarray}

Note that, we use $f^{'}_{a}$ and $\theta^{'}_{a}$ here indicating that the attacker may change $model_{a}$ as well as its parameters specially for $\mathbf{a}^{'}$.

\subsection{Drawbacks of Existing Approach}\label{sec:Problem in Existing Paradigm}
Currently, there are a few research works such as DPFE~\cite{DPFE} and Shredder~\cite{Shredder} focusing on protecting user privacy in the scenario of edge-cloud system running DNN models. However, none of them solves the privacy protection problem entirely.

For DPFE, its privacy protection relies on the well-defined sensitive variables, which means it requires the end user to identify what kind of information they need to protect. However, in practice, the user may not know the exact privacy that should be protected and thus fails to provide such information~\cite{bellekens2016pervasive,malandrino2013privacy}.
Even if such information is provided, the optimization strategy of DPFE makes it privacy specific, which means one DPFE model is only able to protect the specified privacy and has little potential for generalization. The above drawbacks deteriorate the effectiveness of DPFE, and thus prevent its widely adoption in practice.
Whereas for Shredder, although the approach of Shredder is not based on specific privacy information, the metric of Mutual Information (MI) for evaluating the effectiveness of privacy protection targets the entire database rather than a single image, and thus is insufficient to measure the privacy protection for a single user.

Specifically, we argue there are two reasons indicating MI is not a good indicator for evaluating the methods for privacy protection. Firstly, the calculation for MI can only represents \textit{group} privacy rather than \textit{pairwise} privacy. The definition of Mutual Information (MI), $I(X;Y)$ is shown in Eq.~\ref{equ:MI}, where $X,Y$ are discrete random variables in database $\mathcal{D}^{n}$, that is defined on domain $\mathcal{D}$. As illustrated in Shredder, the calculation for MI of $i^{th}$ pixel in the image is denoted as $x_{i}$, and thus the MI for the entire image is denoted as $X=[x_{1} \dots x_{n}] \sim f_{X}$. Similarly for intermediate results, the MI is denoted as $Y=[y_{1} \dots y_{m}] \sim f_{Y}$. Therefore, the calculation of MI is based on estimating the distribution of $X$ and $Y$ over the entire dataset ($f_{X}$ and $f_{Y}$).
Such calculation of MI results in the \textit{group} privacy rather than \textit{pairwise} privacy for a single image~\cite{MI-Relation}. It also means MI is only useful regarding a given test dataset, which fails to represent our scenario of edge-cloud system where the input image from the user cannot be predicted in advance, and thus insufficient to evaluate the effectiveness of privacy protection methods (demonstrated by our experiments in Sec.\ref{sec:Insufficient of MI}). In our scenario, privacy protection should be enforced at \textit{pairwise} level, targeting a single user.

\begin{equation}\label{equ:MI}
    I(X;Y)=\sum_{x,y\in\mathcal{D}^{n}}{p_{X,Y}(x,y)\log\frac{p_{X,Y}(x,y)}{p_{X}(x)p_{Y}(y)}}
\end{equation}

\begin{figure}[htb]
    \setlength{\abovecaptionskip}{1px}
    \begin{center}
       \includegraphics[width=0.9\linewidth]{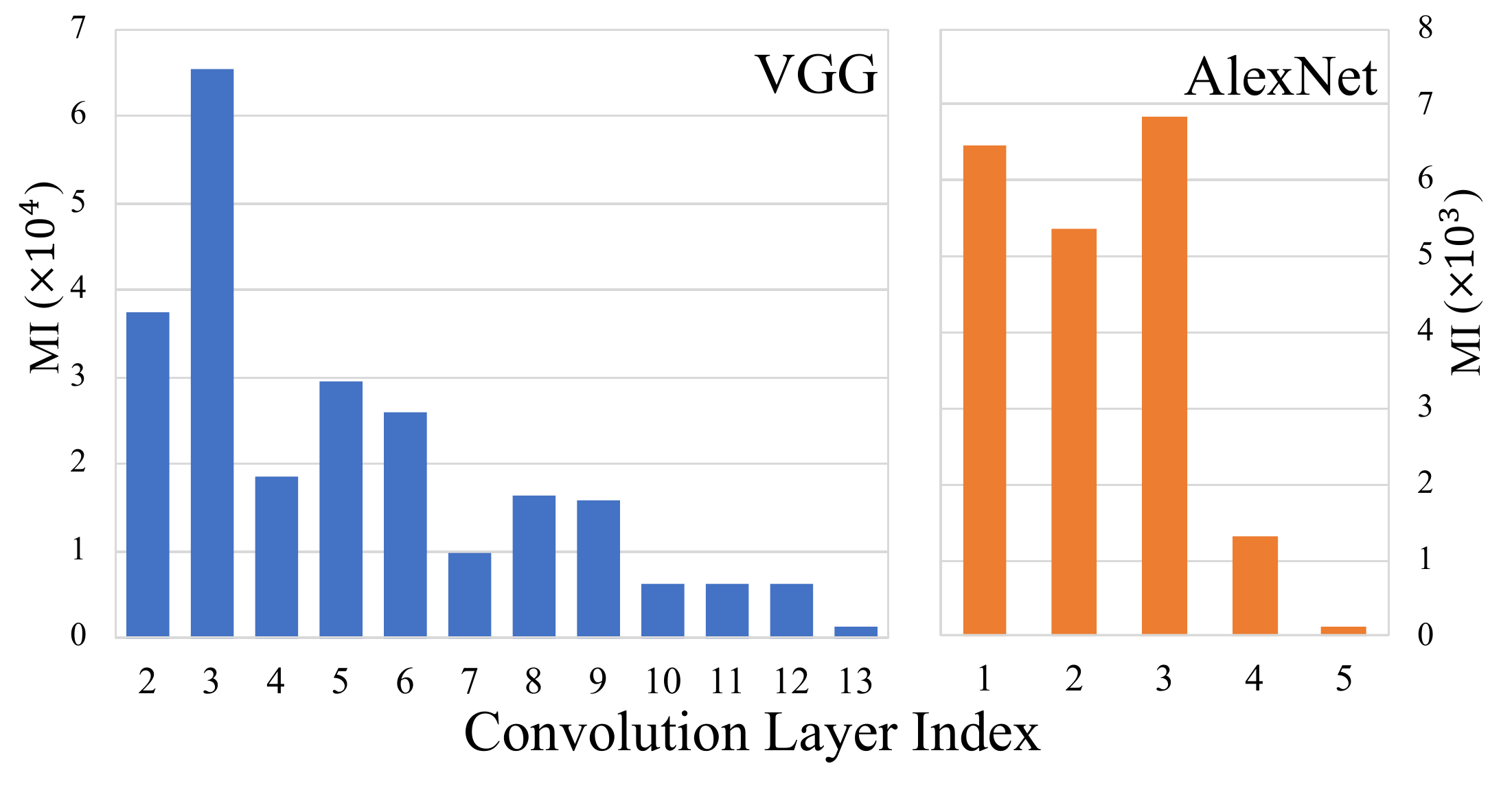}
    \end{center}
    \caption{MI between first channel of original input and intermediate result of after each convolutional layer for VGG16 and AlexNet. The models are fine-tuned on a subset of ML-Image dataset.}
    \label{fig:VGG-alex-MI}
 \end{figure}

Secondly, the calculation of MI dose not accurately reflect the amount of information. 
To prove this observation with experiments, we cut the VGG16 model~\cite{vgg} with BatchNorm layer and the AlexNet model~\cite{Alex} at each convolutional layer.
After fine-tuning the models on a subset of ML-Image dataset~\cite{ml-images}, we calculate (based on the calculation proposed by Shredder) MI between the first channel of original input and the intermediate result.
As shown in Fig.~\ref{fig:VGG-alex-MI}, intermediate results from deeper layers may not lead to a decrease of MI, on the contrary sometimes increase MI.
Based on the MI changes across different layers, we can conclude that information contained at different layers of DNN models fluctuates.
However, deeper layers in a DNN model are considered to provide more abstract representation of the data~\cite{CutDeep}.
According to Data Processing Inequality~\cite{DataProcessingInequality}, information reduces as the depth of the model grows.
Such discrepancy between the MI results and the above principle indicates that MI cannot be used to represent the amount of information accurately.
Therefore, using MI to evaluate the effectiveness of privacy protection applied to intermediate result is insufficient.

Beyond the evaluation metric, the evaluation of both DPFE and Shredder targets much harder tasks for the attacker, which is hardly persuasive for the effectiveness of privacy protection of their approaches. For example, in DPFE the attacker task is defined as identity prediction, which is much harder than the original user task (attributes prediction). Similarly for Shredder, the original user task requires identifying whether the number is greater than 5, however the attacker task is designed to identify the exact number between 0 and 9, which is again a harder task. In such experiment settings, reducing the feature dimension of the intermediate results using the approaches proposed by DPFE and Shredder can deteriorate the accuracy for the attacker task, however whether it is still effective when targeting the attacker task at the same hardness level of the user task remain unclear. Therefore, the existing experiments are insufficient to prove the effectiveness of their methods.


\subsection{A New Testimony}\label{sec:A New Testimony}
In Sec.~\ref{sec:Problem in Existing Paradigm} we have analyzed the drawbacks of existing approaches for privacy protection such as simplified assumption of no model fine-tuning and insufficient evaluation metric MI.
Therefore, in this paper, we formally establish the scenario and propose new metrics to evaluate the effectiveness of privacy protection methods for end user especially on the edge-cloud system running DNN models.
Based on Eq.~\ref{equ:convert function}, the goal is to evaluate the performance of privacy protection. We formulate the scenario for privacy protection as follows. 

\begin{figure}[htb]
    \setlength{\abovecaptionskip}{1px}
    \begin{center}
       \includegraphics[width=0.7\linewidth]{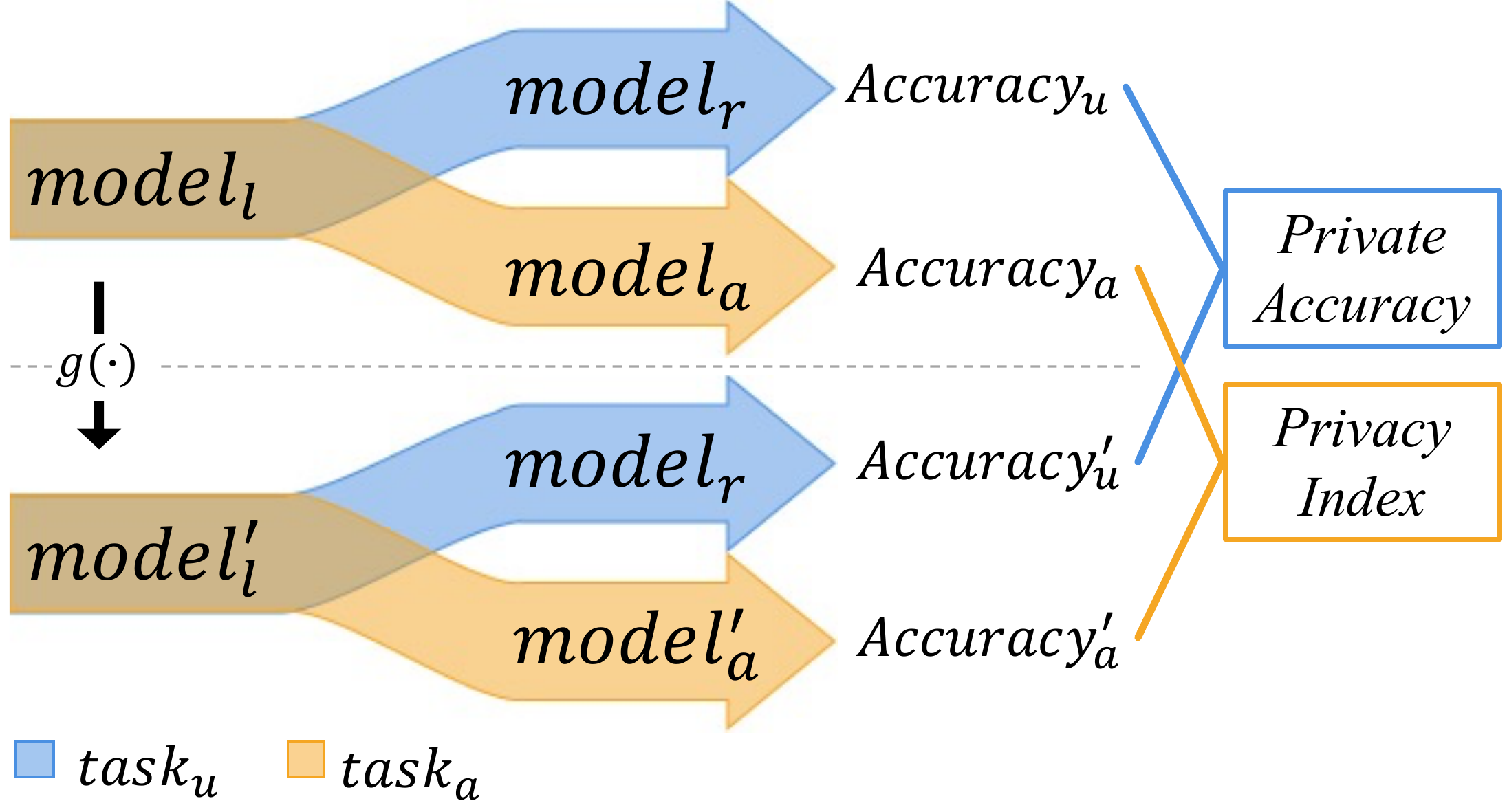}
    \end{center}
    \caption{Relationship between user and attacker models/tasks, with targets for each training procedure.}
    \label{fig:models}
 \end{figure}

We define the \textit{user task} as $task_{u}$.
Based on Assumption~\ref{Ass:question define}, the service provider should only extract information that is useful to generate the correct result for the \textit{user task}, but is prevented from all other information in data $\mathbf{a}^{'}$. Therefore, when designing a privacy protection model $g(\cdot)$, the \textit{attacker task}, $task_{a}$, should be comparable to the $task_{u}$. And a corresponding \textit{attacker model}, $model_{a}=f_{a}(\mathbf{a};\theta_{a})$, should be designed to complete $task_{a}$ very well. After designing the privacy protection model $g(\cdot)$ with regard to $task_{u}$, $model_{l}^{'}=f_{l}^{'}(x;\theta_{l}^{'})$ running on the edge device should be reused to generate input $\mathbf{a^{'}}$ for $task_{a}$.
At this time, a \textit{joint model} is trained that consists of $model_{l}^{'}$ and $model_{a}$, where parameters of $model_{l}^{'}$ are fixed. The \textit{joint model} can be expressed as Eq.~\ref{equ:eq8}.

\begin{equation}\label{equ:eq8}
    \mathbf{y}^{'} = f_{a}(f_{l}^{'}(\mathbf{x}^{'});\theta_{a}^{'})
\end{equation}

Note that, we use $\mathbf{x}^{'},\mathbf{y}^{'}$ here to show that $task_{a}$ may be defined on another dataset, and $\theta_{a}^{'}$ is the only trainable parameter in the \textit{joint model}, where $f_{l}^{'}(\cdot)$ should be considered as a completely specified function.
Here we propose two new metrics \textit{Private Accuracy (PA)} and \textit{Privacy Index (PI)} as shown in Eq.~\ref{equ:PA} and Eq.~\ref{equ:PI}, where $Accuracy_{R}$ refers to accuracy generated by random outputs. Both metrics are within the range of $[0,1]$, with higher value indicates better result. \textit{PA} can be used to evaluate the accuracy loss, whereas \textit{PI} can be used to evaluate the effectiveness of privacy protection. The relationship between \textit{PA} and \textit{PI}, as well as the user and attacker models/tasks is shown in Fig.~\ref{fig:models}. When designing a privacy protection model, there is a special \textbf{constraint}. Although the goal of the method is to generate higher $PI$, one should consider $task_{a}$ as unknown.
Otherwise, $PI$ becomes valid for only $task_{a}$ and loses its potential for generality.

\begin{eqnarray}
    \label{equ:PA}
    PA&=&\frac{Accuracy_{u}^{'}}{Accuracy_{u}} \\
    \label{equ:PI}
    PI&=&\frac{Accuracy_{a} - Accuracy_{a}^{'}}{Accuracy_{a}-Accuracy_{R}}
\end{eqnarray}

In general, our proposed formulation and evaluation metrics have following advantages:
\begin{itemize}
    \item Our scenario and assumption is based on the real-world usage and reveals the significant drawbacks of existing privacy protection approaches, which requires further research efforts to address them.
    \item For edge-cloud system, we formulate the problem for privacy protection when running DNN model on such system, which generalizes future solutions proposed for such scenario.
    \item The new evaluation metrics we proposed can measure the effectiveness of privacy protection at \textit{pairwise} level, targeting each single user, which is more useful than MI that only works for group privacy.
    \item Our formulation and evaluation metrics allow the tradeoff between the strength of privacy protection and model accuracy by adjusting $(model_{a})_{user}$ to different levels of $Accuracy_{a}$, which enables the privacy protection methods more flexible in practical usage. 
    
\end{itemize}

\section{Experiments}\label{sec:experiments}
In this section, we present the experiment results to demonstrate the privacy is vulnerable in the edge-cloud system running DNN models, even with the existing privacy protection approach applied. We choose Shredder~\cite{Shredder} as one of the state-of-the-art methods focusing on privacy protection for evaluation, which is proved to be insufficient for protecting privacy of individual user in our proposed scenario.

In our experiments, we use Open Images from ML-Image~\cite{ml-images}, a dataset specially designed for multi-objective detection, for training and testing the user models. And we use VOC2012~\cite{pascal-voc-2012}, a popular dataset for vision task, for fine-tuning and testing the attacker models.

We define the \textit{user task}, $task_{u}$, as a single-objective classification task on a subset of ML-Image, which contains 11 classes (shop, seabird, generator, bag, seat, needlework, keyboard, trail, basketball player, fungus and salad).
Each class consists of 5,000 images approximately (due to some died image links from the original dataset), which ends up with 53,165 images in total.
We use 48,013 images for training, and 5,152 images for validation (both image sets have the same ratio for each class).
We define the \textit{attacker task} $task_{a}$ as a binary classification task on VOC2012 dataset, identifying whether an image contains human or not. We follow the official image spilt for training and validation.

All models in our experiments are trained using SGD with $momentum=0.9$, and $weight\text{ }decay=3\times10^{-5}$. Besides, we use a label smooth algorithm~\cite{Inception} with $\epsilon=0.1$ for standard $CrossEntropy$ loss. We change the learning rate and apply a decay to learning rate according to different tasks and models for better training results.

\subsection{The Effect of Cutting Point Selection}\label{sec:Best Cutting Point}
In our scenario described in Sec.\ref{sec:Privacy for End User}, we need to cut an existing deep model into two parts, with each part running on the edge device and cloud service respectively. Specifically for a classification model, the cutting point can be applied at different convolutional layers, where privacy protection method should be applied at the cutting point, right after the partial model for feature extraction~\cite{Shredder,DPFE}. According to~\cite{Shredder} the deeper cutting point is, the better the privacy protection is. However, such claim is suspicious, because different models may have different properties, especially considering the factors such as computation and storage requirements that could become overwhelming for edge device. The trade-offs should be evaluated for selecting the cutting point, such as how the selection affects privacy protection and model accuracy. Therefore, we conduct several experiments on AlexNet~\cite{Alex} based on the settings of Shredder (except the dataset, we use ML-Image instead) to measure the effect of cutting point selection.

\begin{figure}[htb]
    \setlength{\abovecaptionskip}{1px}
    \begin{center}
       \includegraphics[width=0.9\linewidth]{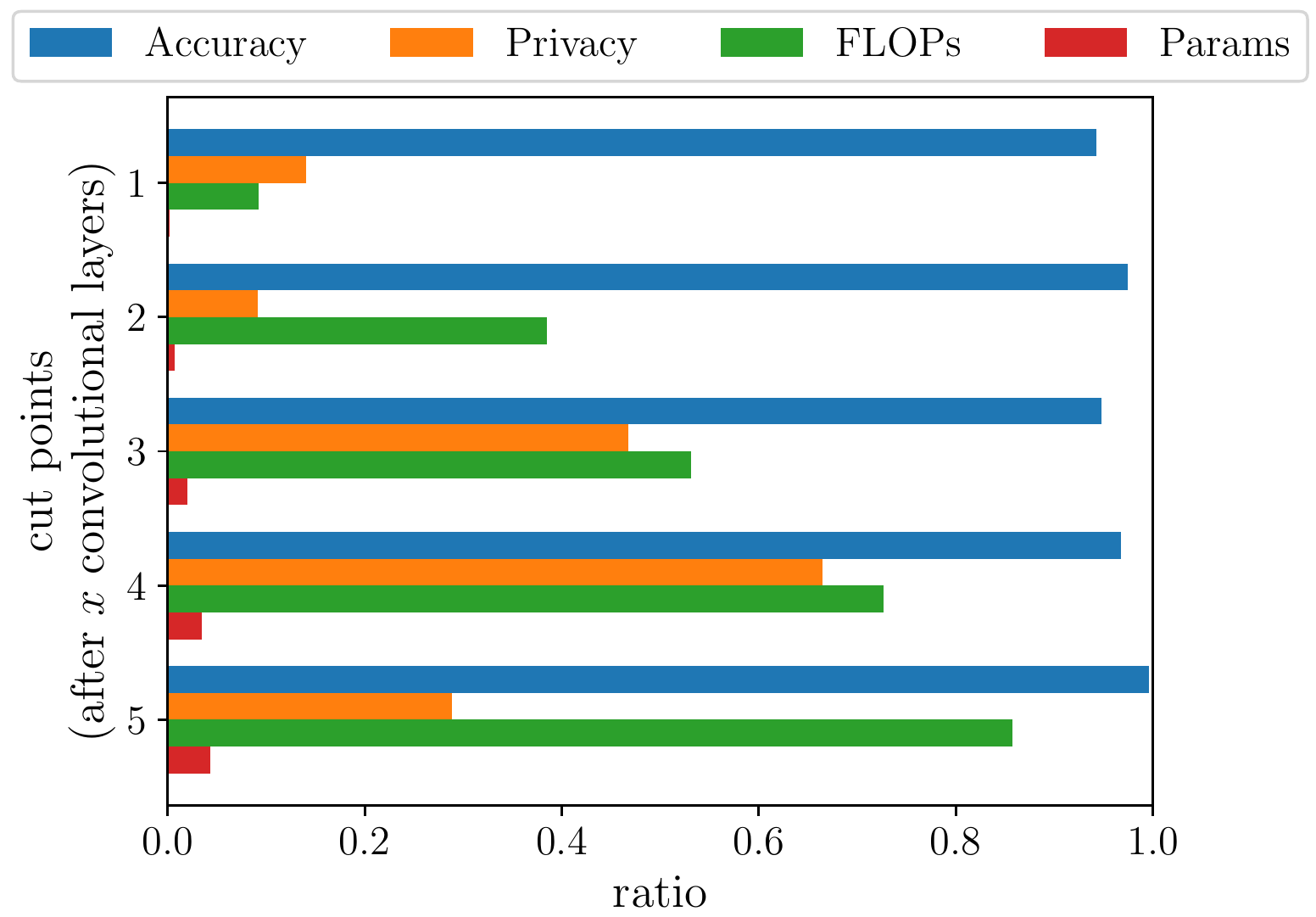}
    \end{center}
    \caption{Performance evaluation of Shredder for different cutting points on AlexNet with ML-Image dataset. Privacy is measured in MI. All metrics are normalized.}
    \label{fig:AlexCutPoint}
\end{figure}

We evaluate the effect of cutting point selection on the deep model from four different aspects, including accuracy and privacy (e.g., MI),  computation (e.g., FLOPs) and storage (e.g., number of parameters).
For the ease of comparison, we normalize all metrics within the range of $[0,1]$.
Accuracy is normalized by the original accuracy on ML-Image using AlexNet ($84.12\%$). MI is normalized based on $\frac{\text{MI}_\text{original}-\text{MI}_\text{noise added}}{\text{MI}_\text{original}}$, which is the higher the better.
FLOPs and Params indicate the ratio of the entire model that is performed on the edge device, which are the lower the better.

As shown in Fig.~\ref{fig:AlexCutPoint}, because most parameters are in the linear layer of AlexNet, the choice of cutting point does not affect the parameters too much. However, since the most computations performed at the convolutional layers, cutting point selection affects the FLOPs significantly. It is seen that if a deeper cutting point is chosen, then the computation pressure is shifted to the edge device. In the extreme case (e.g, at \textit{conv5} layer), almost all computation are performed at the edge device. Regarding the accuracy, adding noise after each of the cutting points does not affect the model accuracy noticeably.
However for MI, there is no clear trend on how the the cutting point affects MI reduction.
Specifically, except \textit{conv4} layer, other selection of cutting point causes a deterioration of privacy protection, which indicates using Shredder for AlexNet, neither too shallow nor too deep layers should be chosen as the cutting points. Such limitation on cutting point selection narrows down the trade-off between computation on edge device and strength of privacy protection.

In the following section, we will focus on the cutting points that achieve strong privacy protection indicated by higher MI reduction, and demonstrate how existing approach becomes invalid when applied in our proposed scenario.

\subsection{Insufficiency of MI}\label{sec:Insufficient of MI}
Given the problem definition in Sec.~\ref{sec:experiments}, if MI is sufficient to represent the effectiveness of privacy protection, we can say that a model with high MI reduction should generate less accurate results on $task_{a}$.
To validate that, we use AlexNet, VGG16\_bn~\cite{vgg} and ResNet50~\cite{resnet} as the base models (for ResNet50, we make the cutting point at the granularity of Bottlenecks to keep its integration), and apply Shredder method to obtain the models that produce intermediate results with decreasing MI. Then we train the \textit{joint model} defined in Sec.\ref{sec:A New Testimony} to determine whether MI is sufficient to indicate the effectiveness of privacy protection.

\begin{figure}[htb]
    \setlength{\abovecaptionskip}{1px}
    \begin{center}
       \includegraphics[width=\linewidth]{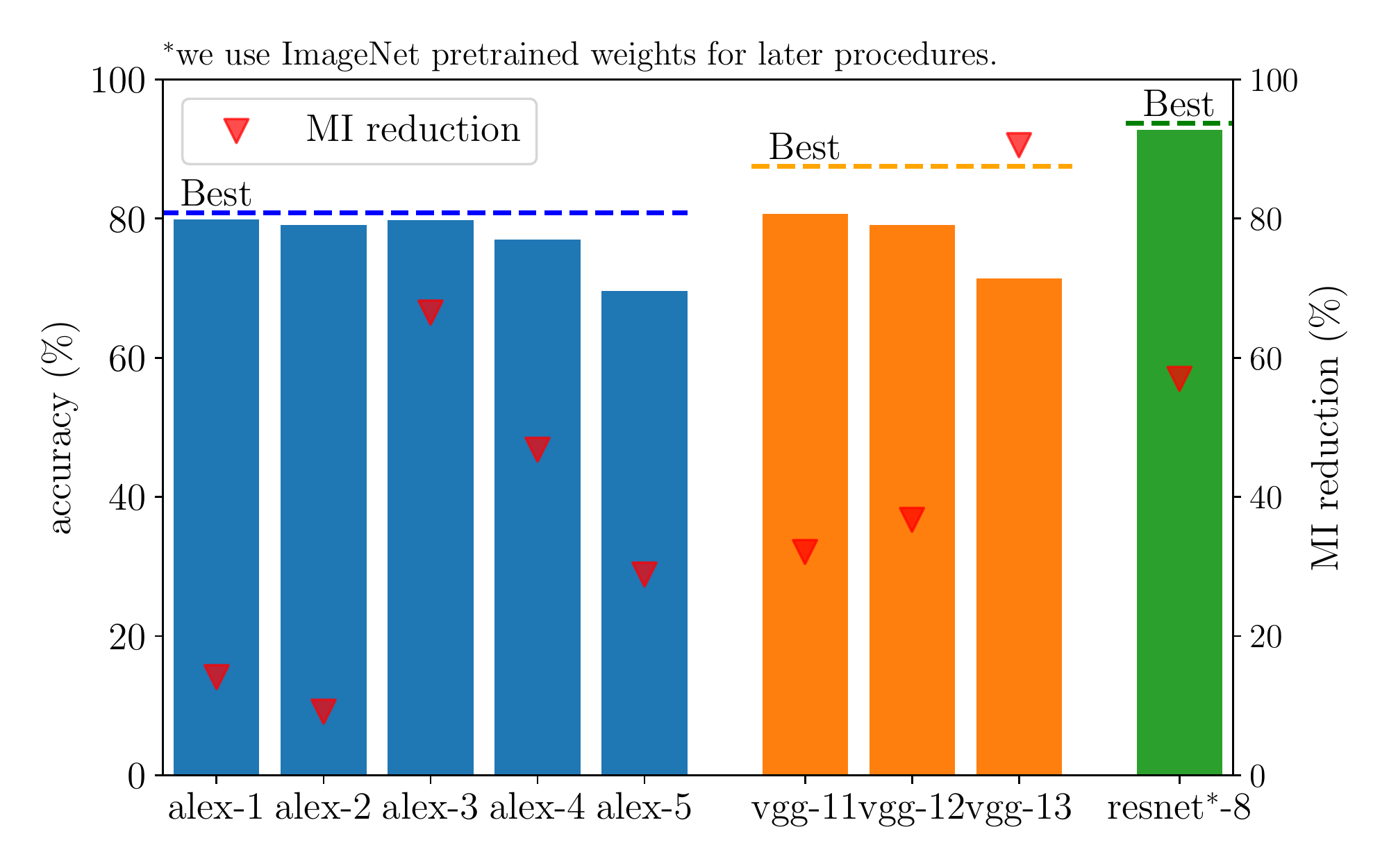}
    \end{center}
    \caption{MI reduction and accuracy for AlexNet, VGG16\_bn and ResNet50 on VOC2012 for $task_{a}$. Dot line indicates the best performance each model can reach without any constraint.}
    \label{fig:Attack}
\end{figure}

As shown in Fig.~\ref{fig:Attack}, the noise applied in the model leads to significant MI decrease compared to the original model.
For the cutting point at \textit{conv13} layer of VGG, the MI reduction even reaches up to 90.54\%. The MI results in Fig.~\ref{fig:Attack} indicate that the privacy of the \textit{user task} is well protected.
However, when the attacker uses another model, $model_{a}$, instead of $model_{r}$ and fine-tune the \textit{joint model} according to Sec.\ref{sec:A New Testimony}, the fine-tuned models can generate high accurate results on $task_{a}$ (slight decrease of accuracy compared to the $Best$ result across various cutting points of different DNNs).
These results indicate that in our scenario, the privacy protection method such as adding noise proposed by Shredder fails to provide privacy protection for the \textit{user task}.
For example, when applying the noise to the cutting point at \textit{conv4} layer of AlexNet, although it achieves the highest MI reduction (e.g., 66.48\%), the model accuracy on $task_{a}$ only reduces by 3.79\% compared to the best accuracy.
Therefore, the above mismatch between MI and model accuracy on $task_{a}$ indicates that MI is insufficient to measure the effectiveness of privacy protection for edge-cloud system.

\begin{table}[htb]
    \small
    \centering
    \begin{threeparttable}
    \begin{tabular}{lccc}  
    \toprule
    Cutting Point  & $PA$ & $PI$ & Accuracy on $task_{a}$ (\%) \\
    \midrule
    AlexNet-1   &	0.94	&	0.03    &   79.86	\\
    AlexNet-2   &	0.97	&	0.05    &   79.10	\\
    AlexNet-3   &	0.95	&	0.03    &   79.80	\\
    AlexNet-4   &	0.97	&	0.12    &   77.00	\\
    AlexNet-5   &	1.00	&	0.36    &   69.64  	\\
    \midrule
    VGG-11      &	0.95	&	0.18    &   80.79	\\
    VGG-12      &	0.97	&	0.21    &   79.57	\\
    VGG-13      &	0.98	&	0.43    &   71.45   \\
    \midrule
    ResNet\tnote{*}-8   &   0.97    &   0.02 &   92.76 \\
    \bottomrule
    \end{tabular}
        \begin{tablenotes}
            \footnotesize
            \item[*] we use ImageNet pre-trained weights for later procedures.
        \end{tablenotes}
    \end{threeparttable}
    \caption{\small$PA$ and $PI$ are calculated for AlexNet, VGG16\_bn and ResNet50 with different cutting points. (8$^{th}$ layer is the deepest cutting point where the FLOPs required on edge device are smaller than on cloud for ResNet50.)}
    \label{tab:Attack}
\end{table}

On the contrary, $PA$ and $PI$ proposed in this paper are better metrics for evaluating the effectiveness of privacy protection, as shown in Tab.~\ref{tab:Attack}.
Higher $PA$ means better accuracy for the \textit{user task}, whereas higher $PI$ means lower accuracy for the \textit{attacker task} and thus better protection for privacy.
Different from MI, $PI$ can accurately reflect the accuracy of $task_{a}$, with higher $PI$ indicating better privacy protection.
For example, in Tab.~\ref{tab:Attack}, $PI$ indicates \textit{conv5}, \textit{conv13} and \textit{conv8} can achieve the best privacy protection among all the cutting points for AlexNet, VGG16\_bn and ResNet50 respectively, which is consistent with the least accuracy of $task_{a}$. Therefore, $PI$ is a better metric for evaluating the effectiveness of privacy protection methods for edge-cloud system.

\section{Discussion}
In this paper, we formulate the privacy protection problem for DNN models running on edge-cloud system, revisit the assumption for privacy breach and advocate two new metrics for evaluating the effectiveness of privacy protection methods. We investigate the existing approaches with experiments to demonstrate their drawbacks in the determining the cutting point for privacy protection, as well as show the consistency of our new metrics in measuring the accuracy of \textit{user task} and the level of privacy protection.

This paper serves as the position paper that calls for research efforts on privacy protection for DNN models running on edge-cloud system. We highlight several take-aways as follows to encourage future research efforts on this specific research direction.


\textbf{Take-away 1: Dynamic noise regarding each input image is required to protect intermediate results}. Adding noise has been proved effective to restrict the information contained in intermediate results. However, existing methods are designed without awareness of inputs. The noise used in such methods fails to consider the specific features of input image, which also limits their usage of privacy protection for a single user, or \textit{pairwise} privacy.
Therefore, using input feature to generate noise can be one potential direction to achieve \textit{pairwise} privacy.

\textbf{Take-away 2: The noise can be added at multiple cutting points simultaneously}.
Currently, existing approaches only consider adding the noise at a single cutting point of the DNN models. Since the noise will not generate too much computation cost on the edge device, adding noise at multiple cutting points can be another promising solution to further restrict the information that can be extracted from the intermediate results by the attacker.

\textbf{Take-away 3: New optimization functions should be designed regarding the new evaluation metrics}.
In this paper, we propose new evaluation metrics $PA$ and $PI$.
However, there are no optimization functions currently available for them. Especially for $PI$, the optimization function should also consider the constraints described in Sec.~\ref{sec:A New Testimony}. Therefore, new optimization functions should be designed regarding $PA$ and $PI$ in order to achieve effective privacy protection.

We hope this paper can shed the light in the research direction of privacy protection for DNN models running on edge-cloud system. We wish the discussion highlights can pave the road for future research and encourage more researchers to work in this field.

\bibliographystyle{named}
\bibliography{main}

\end{document}